\documentclass{article}
\usepackage{ijcai21}

\usepackage{times}
\usepackage{soul}
\usepackage{url}
\usepackage[hidelinks]{hyperref}
\usepackage[utf8]{inputenc}
\usepackage[small]{caption}
\usepackage{graphicx}
\usepackage{amsmath}
\usepackage{amsthm}
\usepackage{booktabs}
\usepackage{algorithm}
\usepackage{algorithmic}
\usepackage{multirow}
\usepackage{amsfonts}
\title{Exploring Periodicity and Interactivity in 
Multi-Interest Framework for Sequential Recommendation}

\author{
Gaode Chen$^{1,2}$
\and
Xinghua Zhang$^{1,2}$\and
Yanyan Zhao$^{1,2}$\and
Cong Xue$^{1}$\footnote{Cong Xue is the corresponding author.}\And
Ji Xiang$^{1,2}$
\affiliations
$^1$Institute of Information Engineering, Chinese Academy of Sciences, Beijing, China\\
$^2$School of Cyber Security, University of Chinese Academy of Sciences, Beijing, China
\emails
\{chengaode, zhangxinghua, zhaoyanyan, xuecong, xiangji\}@iie.ac.cn
}

\begin{document}

\maketitle

\begin{abstract}
Sequential recommendation systems alleviate the problem 
of information overload, and have attracted increasing 
attention in the literature. Most prior works usually 
obtain an overall representation based on the user's 
behavior sequence, which can not sufficiently reflect 
the multiple interests of the user. To this end, we 
propose a novel method called PIMI to mitigate this 
issue. PIMI can model the user's multi-interest 
representation effectively by considering both the periodicity and interactivity in the item sequence. 
Specifically, we design a periodicity-aware module to 
utilize the time interval information between user's 
behaviors. Meanwhile, an ingenious graph is proposed to 
enhance the interactivity between items in user's behavior 
sequence, which can capture both global and local item 
features. Finally, a multi-interest extraction module is applied to describe user's multiple interests based on the obtained item representation. Extensive experiments on two real-world 
datasets Amazon and Taobao show that PIMI outperforms 
state-of-the-art methods consistently. 
\end{abstract}

\section{Introduction}

Sequential recommendation systems play an important role in helping users alleviate the problem of information overload, and 
in many application domains, e.g., ecommerce, social 
media and music, it can help optimize the business metrics 
such as click-through rate (CTR). Sequential recommendation systems 
sort items by the timestamp of user behavior, 
and focus on sequential pattern mining to predict the next 
item that users may be interested in. Most existing 
methods combine user's preference and item representation 
to make predictions, researches in sequential 
recommendation are therefore largely concerned with how 
to improve the representation quality of users and items. 

Due to sequential recommendation systems' highly practical value, 
many kinds of approaches for sequential recommendation have
been proposed and achieved promising performance. For example, GRU4Rec \cite{hidasi2015session} is the first work to apply RNN to model the sequence information for recommendation. 
~\citeauthor{kang2018self}~\shortcite{kang2018self} 
propose attention-based method to capture high-order dynamic information in the sequence. 
Recently, some works (e.g. PinSage~\cite{ying2018graph}) 
leverage Graph Neural Network (GNN) based methods to 
obtain the representation of users and items for 
downstream tasks. 
However, we observe that most prior studies 
obtain an overall representation for user's behavior 
sequence, but a unified user embedding is difficult to 
reflect the user's multiple interests. 
\begin{figure}[t]
\centering
\includegraphics[width=0.49\textwidth,trim=20 190 0 0,clip]{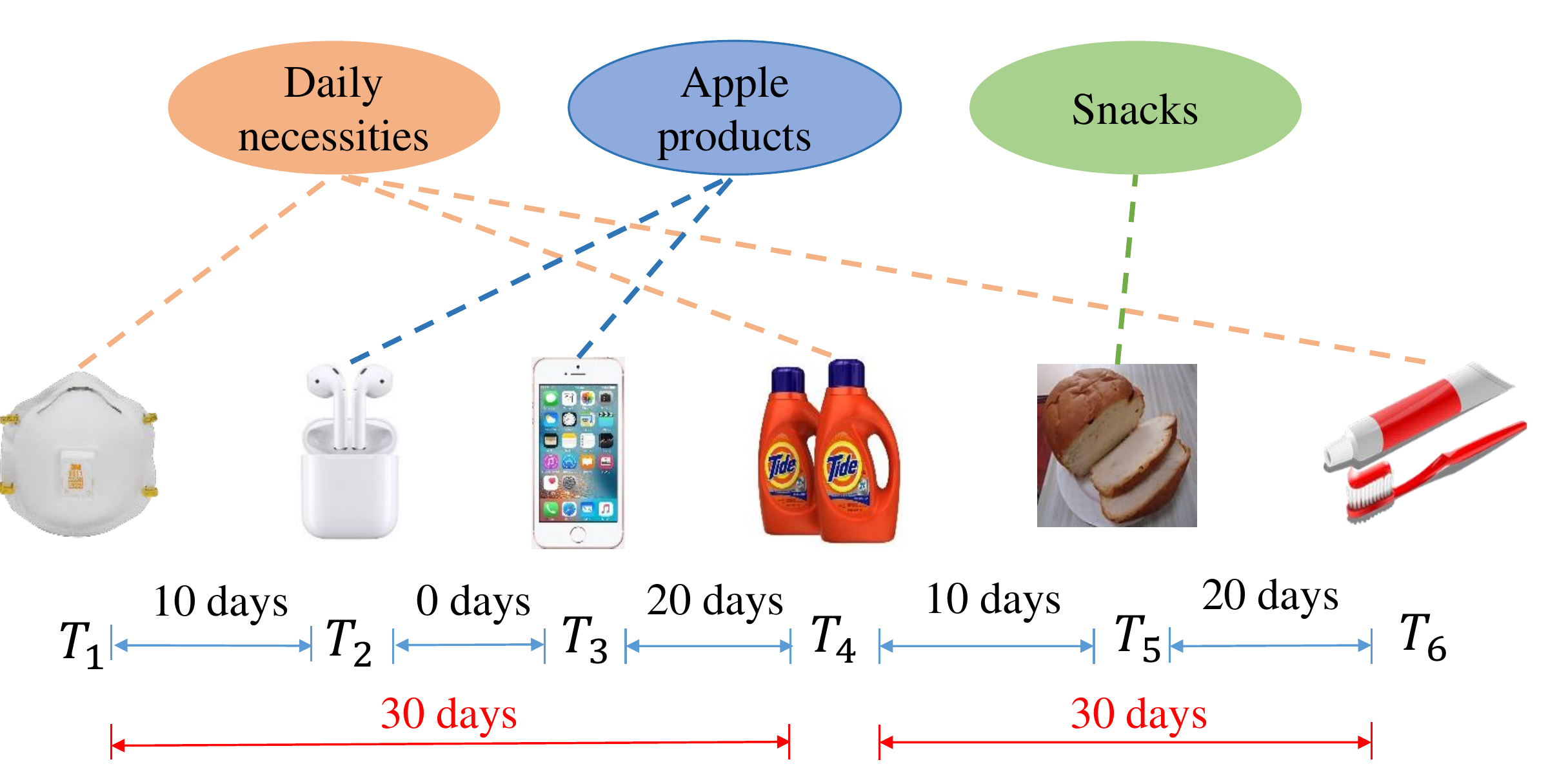}
\caption{The behavior of different interests have different time 
periodicity in user sequence.}
\label{Figure 1}
\end{figure}
In the literature, few studies attempt to model the 
multi-interest representation of users to alleviate the 
problem of insufficient represent ability of a single 
vector. 

Recently, MIND \cite{li2019multi} utilizes a 
dynamic routing method based on the capsule network 
\cite{sabour2017dynamic} to 
adaptively aggregate user's historical behavior into 
user's multiple representation vectors, which can reflect the different 
interests of the user. ComiRec \cite{cen2020controllable} 
leverages self-attention mechanism and dynamic routing 
method for multi-interest extraction following MIND. 
However, these methods have the following limitations: 
(1) They only use time information to sort 
items, and ignore that the behaviors of different 
interests have different time periodicity in user sequence. 
For example, in Figure 1, given a user’s behavior 
sequence, the user may be interested in daily necessities, 
Apple's products and snacks. He/she may buy daily 
necessities every month, but he/she only pays attention 
to Apple's products during the launch of new Apple products.  
Therefore, the time interval for the interest in daily 
necessities is about one month, while the time interval 
for the interest in Apple's products is longer, 
about one year. In summary, 
users’ behavior for different types of items have 
different periodicities. 
(2) The interactivity between items is not explored 
effectively. These methods only model 
the correlation 
between adjacent items in the sequence, but do not 
consider the interactivity 
between items in the multi-interest extraction. 
In fact, multi-interest extraction can be viewed as the process of soft clustering between items, and the interactivity 
between items is effective for clustering tasks 
\cite{zhang2017learning}, because items of the same 
type will learn similar representation through 
interaction. 
Thus, we argue that the time interval and 
interaction information between items in the user's sequence are more powerful to capture multi-interest representation. 

To solve these problems, we propose a novel method, called 
PIMI, to explore \textbf{P}eriodicity and 
\textbf{I}nteractivity in \textbf{M}ulti-\textbf{I}nterest 
framework for sequential recommendation. Firstly, we encode the time interval information between items in the sequence so 
that the periodicity information can be involved in the user's 
multi-interest representation, which can reflect the 
dynamic changes of the user behavior. Secondly, 
we design an ingenious graph structure. 
Previous GNN-based methods ignore the sequential
information in the user behavior, our graph structure 
overcomes this shortcoming and captures the
correlation between adjacent user behavior.  
What's more, the proposed graph structure can gather and scatter 
global and local item interactivity information with the 
virtual central node to improve the performance of 
multi-interest extraction. Finally, we obtain 
multi-interest representation for user based on the 
attention mechanism, which can be used to select candidate items and make recommendations. 
The main contributions of this work are summarized as 
follows: 
\begin{itemize}
\item We incorporate the time interval information in the 
user behavior sequence, which can model the periodicity 
of user's multiple interests and improve the quality of 
the user’s representation. 
\item We design an innovative graph structure to capture 
the global and local interactivity among items, and 
retain the sequential information at the same time, which can 
improve the quality of the item's representation.
\item Our model PIMI achieves the state-of-the-art performance on
two real-world challenging datasets Amazon and Taobao 
for the sequential recommendation.
\end{itemize}

\section{Related Work}
\paragraph{Sequential recommendation.} 
Sequential recommendation systems are based on the user's 
behavior  
sequence to predict the next item that the user might be 
interested in. Many recent works about sequential 
recommendation focus on this problem. 
FPMC \cite{rendle2010factorizing} contains a common Markov chain and the normal 
matrix factorization model for sequential data.  
SDM \cite{lv2019sdm} combines user's long- and short-term 
preferences to make recommendations, which models user's 
preferences based on LSTM and dense fully-connected 
networks. 
Chorus \cite{wang2020make} utilizes the information of 
the knowledge graph to model the relations between 
items, and introduces temporal kernel functions 
for each item relation to better capture dynamic user demands. 
These methods give a single vector representation of the 
user based on behavior sequence, which is hard to 
reflect the real-world recommendation situation. 
Recently, MIND \cite{li2019multi} and ComiRec \cite{cen2020controllable} 
attempt to use dynamic routing-based methods and 
attention-based methods  
to obtain multiple user's vectors to reflect multiple 
interests in the sequence.
However, they do not explore the periodicity of multiple 
interests and the interactivity between items sufficiently, 
which are conducive to the extraction of multi-interest.

\paragraph{Time information learning for recommendation.} 
Time information is very important for 
recommendation. 
Most sequential recommendation methods sort items 
according to the timestamp of user's interactions, 
which implicitly uses time information. 
Few works attempt to model the time information 
in the sequence explicitly. 
For example, MTIN \cite{jiang2020aspect} develops a 
parallel temporal mask network, 
which is able to learn multiple temporal information for 
recommendation. 
TiSASRec \cite{li2020time} combines the advantages of 
absolute position and relative
time interval encodings based on self-attention to predict 
future items.  

\paragraph{Graph neural network.} 
Graph embedding is to learn a mapping function which maps
the nodes in a graph to low-dimensional latent ]
representation 
\cite{zhou2018graph}. 
Some recent works utilize graph neural network \cite{scarselli2008graph} 
methods to 
obtain the representation of users and items, 
which can be used for recommendation tasks. For example, 
GATNE \cite{cen2019representation} supports both 
transductive and inductive embedding learning for 
attributed multiplex heterogeneous networks, which can 
learn the representation of users and items. 
However, GNNs essentially deal with the interactivity 
between nodes, they neglect the relationship 
between adjacent items in the user sequence.  

\paragraph{Attention.} 
The originality of attention mechanism can be traced
back to decades ago in fields of computer vision 
\cite{xu2015show}.  
It is also adapted
to recommendation systems and rather useful on real-world
recommendation tasks. 
For instance, SASRec \cite{kang2018self} captures 
high-order dynamics in user behavior sequences based on 
the self-attention mechanism. 
GC-SAN \cite{xu2019graph} designs a multi-layer 
self-attention network 
to obtain contextualized non-local representation in 
sequence. 
CoSAN \cite{2020Collaborative} proposes the collaborative self-attention network to
learn the session representation by modeling the long-range
dependencies between collaborative items. 

\section{Our Method}
The existing sequential recommendation methods usually use a single vector to represent the user, it is hard 
to reflect user's multiple interests in real-world. 
Based on the above observation, we explore using multiple 
vectors to represent the user's multiple interests. 

The recent multi-interest frameworks for sequential 
recommendation ignore two problems: the periodicity of 
user's interest and the interactivity between items in 
the sequence. We believe that the user’s points of 
interest have different time periodicity, so adding time 
information into the multi-interest vectors improves 
the quality of user representation. Meanwhile, the 
interactivity 
between items in the sequence are effective 
in improving the quality of item representation. 

In this section, we first formalize the sequential recommendation 
problem. Then, we propose a novel method, to explore 
\textbf{P}eriodicity and \textbf{I}nteractivity in 
\textbf{M}ulti-\textbf{I}nterest framework 
for recommendation, called PIMI (as shown in Figure 2).

\begin{figure}[t]
\centering
\includegraphics[width=8.6cm]{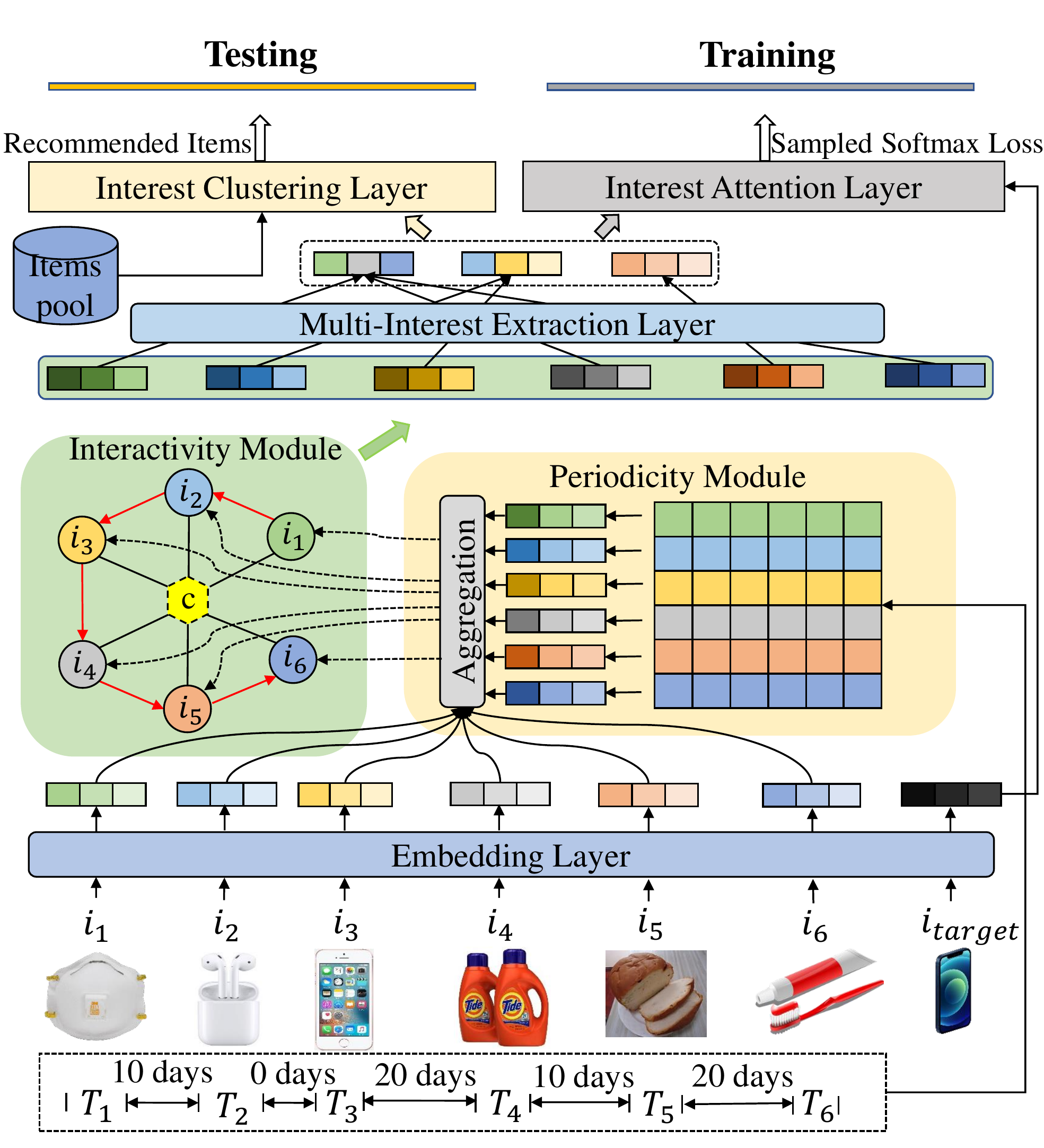}
\caption{The architecture of our proposed PIMI. 
The input of PIMI is the user's behavior sequence, 
which contains a series of item IDs. The item IDs are fed 
into the embedding layer to get the embedding representation 
of each item. Meanwhile, we construct a time interval 
matrix in the periodicity module and introduce the time 
information. We 
aggregate the embedding and time interval representation 
of each item and feed them into the interactivity module.
The items' feature learned from the interactivity module 
are used to generate the user's multi-interest representation 
in the multi-interest extraction layer.
Further more, the multi-interest vectors can be used to compute the 
sampled softmax loss through 
interest attention layer for training, or extract 
candidate items from the global items pool through interest 
clustering layer for testing. 
}
\label{Figure 2}
\end{figure}

\subsection{Problem Statement}
Assume $U$ denote a set of users and 
$I$ denote all items. Each user $u \in U$ 
has a sequence of historical behavior in order. Given 
the user action history $S^u$ 
= $(i_1^u, i_2^u, ..., 
i_{|{S^u}|-1}^u)$, 
$i_r^u \in I$ represents the 
$r$-th item in the sequence. The goal of 
sequential recommendation is to predict the next item that 
the user might be interested in. 

\subsection{Multi-Interest Framework}
\subsubsection{Embedding Layer}
As shown in Figure 2, the input of PIMI is a user behavior
sequence, which contains a series of item IDs 
representing the user’s actions with items in time 
order. 
We convert the user behavior 
sequence ($i_1^u$, $i_2^u$, ..., 
$i_{|{S^u}|-1}^u$) into a 
fixed-length sequence $s^u$ = ($i_1^u$, $i_2^u$, ..., 
$i_n^u$), where $n$ represents the maximum sequence 
length we consider. If the sequence length is greater 
than $n$, 
we truncate it and only take the most recent $n$ items, 
otherwise, we pad the sequence to a fixed length $n$. 

We construct an embedding matrix $M^I \in \mathbb{R}^{|I| 
\times d}$ for all items, where $d$ is the dimension of 
embedding vector. The embedding look-up 
operation converts the IDs of items in the sequence 
into a unified low-dimension latent space. We can obtain 
its embedding: 
\begin{align}
    E^I = [e_1, ..., e_n] \in \mathbb{R}^{n \times d}
\end{align}
where $e_r \in 
\mathbb{R}^{1 \times d}$ is the embedding of the $r$-th item. 

\subsubsection{Periodicity Module} 
Corresponding to the user's behavior sequence $s^u$, 
we can also obtain a time sequence $t^u$ = ($t_1^u$, 
$t_2^u$, ..., $t_n^u$), which 
contains the timestamp of each item in order. 
We only focus on the relative length of the time interval 
in one user behavior sequence and model it as the 
relationship between any two items. Specifically, given 
a fixed-length time sequence $t^u$ = ($t_1^u$, 
$t_2^u$, ..., $t_n^u$) of user $u$, the time interval 
$d_{ab}^u$ between 
item $a$ and item $b$ is defined as the number of days interacted by user $u$, where 
$d_{ab}^u \in \mathbb{N}$. $d_{ab}^u$ and $d_{ba}^u$ are 
equal according to this definition. 
We also set a threshold $p$ 
for the time interval to avoid sparse encoding, 
$d_{ab}^u = min(p, d_{ab}^u)$. Hence, the time interval matrix 
$M^t \in \mathbb{N}^{n \times n}$of a user sequence is: 
\begin{align}
    M^t = \begin{bmatrix} d^u_{11} & d^u_{12} & ... & d^u_{1n} \\ d^u_{21} & d^u_{22} & ... & d^u_{2n} \\ ... & ... & ... & ... \\ d^u_{n1} & d^u_{n2} & ... & d^u_{nn} \end{bmatrix}
\end{align}

Similar to the items' embedding, time interval embedding 
matrix is $M^T \in \mathbb{R}^{n \times n \times d}$. 
For each item in the sequence, we use the time-aware attention  
method to obtain the attention score matrix 
$A_1 \in \mathbb{R}^{n \times n}$ 
of the time interval matrix: 
\begin{align}
    A_1 = softmax(M^TW_1)^\top
\end{align}
where $W_1 \in \mathbb{R}^{d}$ is a trainable parameter. 
The superscript $\top$ denotes the transpose of the 
matrix. The attention score matrix $A_1$ with size $n \times n$ represents 
the attention weight of each item for the time interval of other 
items in the sequence. When we sum up the embedding of 
time intervals according to the attention score, 
the broadcast mechanism in Python is used here, 
we can obtain a matrix representation $E^T \in \mathbb{R}^{n 
\times d}$ of items, which denotes the position of each item 
in the timeline of the overall sequence: 
\begin{align}
    E^T = A_1M^T
\end{align}

\subsubsection{Interactivity Module} 
After the embedding layer and the periodicity module, 
we aggregate the embedding $E^I$ and time interval 
representation $E^T$ of the items and feed them into the 
interactivity module. In the interactivity module, 
we design an ingenious graph structure that regards each item in 
the sequence as a node. Our graph structure not only 
captures sequential 
information, but also allows items to interact via the graph neural 
network. Experimental results prove that the interaction 
between items can effectively improve multi-interest soft 
clustering.

Firstly, we construct a meaningful graph from 
the sequence. As shown in Figure 2, the graph structure 
contains one virtual central 
node and $n$ item nodes. 
The virtual central node is responsible 
for receiving and distributing feature among all item nodes. 
For each item node, the black edge represents the 
undirected connection with the virtual central node. 
Such a graph structure can make any two non-adjacent item 
nodes become two-hop neighbors, and can capture non-local 
information. Since the user's behavior is a 
sequence, we connect the item node in order, as 
shown by the red connection in the graph. Such 
graph structure can model the correlation between adjacent 
items, allow each item node to gather information from 
neighbors, and capture local information. 

Next, we present how to obtain feature vectors of 
nodes via graph neural network. We use $c^l \in \mathbb{R}^{1 
\times d}$ and $H^l \in \mathbb{R}^{n \times d}$ to represent  
the virtual central node and all the item nodes at 
step $l$ respectively. We initalize $H^0$ and $c^0$ as: 
\begin{align}
    H^0 = E^I + E^T
\end{align}
\begin{align}
    c^0 = average(H^0)
\end{align}

The update of all nodes at step $l$ is divided into two stages: 
updating all item nodes and updating the virtual central node. 

In the first stage, each item node aggregates 
the following information: 
its adjacent node $h^{l-1}_{r-1}$ in sequence for local information, 
and the virtual central node $c^{l-1}$ for global information, 
in addition, its previous feature $h^{l-1}_r$, 
and its corresponding item embedding $e_r$. 
After that, we update the feature of each item node $r$ at step 
$l$ based on the attention mechanism.
\begin{align}
    g^l_r = concat[h^{l-1}_{r-1};c^{l-1};h^{l-1}_r;e_r]
\end{align}
\begin{align}
    h^l_r = MultiAtt(Q=h^{l-1}_r, K=g^l_r, V=g^l_r)
\end{align}
where $MultiAtt$ means Multi-Head Attention network. It 
was proposed by ~\citeauthor{vaswani2017attention}~\shortcite{vaswani2017attention}.

In the second stage, the virtual central node aggregates the 
information of all the item nodes $H^l$ and its previous 
feature $c^{l-1}$. 
Similar to the item node, it also uses the attention mechanism 
to update the state.
\begin{align}
    q^l = concat[c^{l-1};H^l]
\end{align}
\begin{align}
    c^l = MultiAtt(Q=c^{l-1}, K=q^l, V=q^l)
\end{align}

The overall update algorithm of the interactivity 
module is shown in the Alg-1.

After $L$ rounds of update, the final feature matrix $H^l \in \mathbb{R}^{n \times d}$ of item nodes 
can be used for multi-interest extraction of user interaction 
sequence.

\begin{algorithm}[tb]
\caption{Update Procedure}
\label{alg:algorithm}
\textbf{Input}: Items embedding $E^I$ and time interval embedding $E^T$ \\
\textbf{Parameter}: Multi-head attention network parameters \\
\textbf{Output}: The feature representation $H \in \mathbb{R}^{n \times d}$ of all items
\begin{algorithmic}[1] 
\STATE $H^0 \gets E^I + E^T$ \\
\STATE $c^0 \gets average(H^0)$ \\
\FOR {$step \gets 1\ to\ L$}
    \FOR {$r \gets 1\ to\ n$}
            \STATE $g^l_r = concat[h^{l-1}_{r-1};c^{l-1};h^{l-1}_r;e_r]$ \\
            \STATE $h^l_r = MultiAtt(Q=h^{l-1}_r, K=g^l_r, V=g^l_r)$ \\
    \ENDFOR
    \STATE $q^l = concat[c^{l-1};H^l]$ \\ 
    \STATE $c^l = MultiAtt(Q=c^{l-1}, K=q^l, V=q^l)$ \\
\ENDFOR
\end{algorithmic}
\end{algorithm}

\subsubsection{Multi-Interest Extraction Layer} 
We use the self-attention method to extract multi-interest 
from the user sequence. Given the hidden embedding 
representation 
$H \in \mathbb{R}^{n \times d}$ of all items from 
the interactivity module. 
We can obtain the attention weight $A_2 \in \mathbb{R}^{K 
\times n}$ of multi-interest by the formula: 
\begin{align}
    A_2 = softmax(W_3tanh(W_2H^\top))
\end{align}
where $W_3$ and $W_2$ are trainable parameters of size 
$K \times 4d$ and $4d \times d$. $K$ denotes the number of user interests. The matrix $A_2$ with size $K \times n$ 
represents the $K$ perspectives of the user sequence, reflecting 
the $K$ interest of the user $u$. Hence, the weighted 
sum of all item embedding with the attention 
weight can obtain the $K$ vector representation of the 
user to reflect the different interests.
\begin{align}
    M_u = A_2H
\end{align}

\subsection{Training Phase} 
After computing the interest embedding from
user behavior through the multi-interest extraction layer, 
based on a hard 
attention strategy in the interest attention layer, for 
the target item, we use the 
$argmax$ operation to find the most relevant one among 
the $K$ vector representation: 
\begin{align}
    m_u = M_u[:, argmax(M_u^\top e_o)]
\end{align}
where $M_u$ is user's multi-interest representation matrix, 
$e_o$ represents the embedding of the target item.

Given a training sample ($u$, $o$) with the user 
embedding $m_u$ and the target item embedding $e_o$, we 
should maximize the probability of user $u$ interacting 
with item $o$ in the training phase. 
Due to the expensive computational cost, we utilize the 
sample softmax method to calculate the likelihood of the 
user $u$ interacting with the target item $o$. Finally, 
we train our model by minimizing the following objective 
function: 
\begin{align}
    \mathcal{L}(\theta)=\sum_{u \in U} -\log \frac{exp(m_u^\top e_o)}{\sum_{v \in Sample(I)} exp(m_u^\top e_v)}
\end{align}

\subsection{Testing Phase}
After the multi-interest extraction layer, we obtain 
multiple interests
embedding for each user based on his/her past behavior,  
which can be used for recommendation prediction. 
In the testing phase, each interest embedding 
can independently cluster top $N$ items 
from global items pool based on the inner product 
similarity 
by the nearest neighbor library such as Faiss 
\cite{johnson2019billion} in the interest clustering layer. 
Hence,we can obtain 
$K \times N$ candidate items, and then get the final recommendation 
results by maximizing the following value function, 
that is, 
a set $R$ containing $N$ items:
\begin{align}
    Q(u, R) = \sum_{x \in R} \max \limits_{1\leq k \leq K} (e^\top_xm^k_u)
\end{align}
where $e_x$ is the embedding of the candidate item, $m^k_u$ 
denotes the $k$-th interest embedding of the user $u$.

\section{Experiments}

\begin{table}[t]
\centering
\Huge
\resizebox{\linewidth}{!}{
\begin{tabular}{ccccc}
\toprule
Dataset      & users & items   & interactions & avg time interval \\ \midrule
Amazon Books & 459,133 & 313,966   & 8,898,041  & 76 days   \\
Taobao       & 976,779 & 1,708,530 & 85,383,796    & 1 day       \\ \bottomrule
\end{tabular}}
\caption*{Table 1: statistics of datasets.}
\label{tab:booktabs}
\end{table}

\begin{table*}[t]
\Huge
\resizebox{\linewidth}{!}{
\begin{tabular}{lcccccccccccc}
\toprule
\multicolumn{1}{c}{\multirow{3}{*}{}} & \multicolumn{6}{c}{Amazon Books}                                & \multicolumn{6}{c}{Taobao}                                      \\ \cmidrule(r){2-7} \cmidrule(r){8-13}
\multicolumn{1}{c}{}                                & \multicolumn{3}{c}{Metrics@20} & \multicolumn{3}{c}{Metrics@50} & \multicolumn{3}{c}{Metrics@20} & \multicolumn{3}{c}{Metrics@50} \\ \cmidrule(r){2-4} \cmidrule(r){5-7} \cmidrule(r){8-10} \cmidrule(r){11-13}
\multicolumn{1}{c}{}                                & Recall   & NDCG    & Hit Rate  & Recall   & NDCG    & Hit Rate  & Recall   & NDCG    & Hit Rate  & Recall   & NDCG    & Hit Rate  \\ \midrule
\textbf{GRU4Rec} \cite{hidasi2015session}            & 4.057    & 6.803   & 8.945     & 6.501    & 10.369  & 13.666    & 5.884    & 22.095  & 35.745    & 8.494    & 29.396  & 46.068    \\
\textbf{YouTube DNN} \cite{covington2016deep}        & 4.567    & 7.670   & 10.285    & 7.312    & 12.075  & 15.894    & 4.205    & 14.511  & 28.785    & 6.172    & 20.248  & 39.108    \\
\textbf{MIND} \cite{li2019multi}                & 4.862    & 7.933   & 10.618    & 7.638    & 12.230  & 16.145    & 6.281    & 20.394  & 38.119    & 8.155    & 25.069  & 45.846    \\
\textbf{ComiRec-DR} \cite{cen2020controllable}         & 5.311    & 9.185   & 12.005    & 8.106    & 13.520  & 17.583    & 6.890    & 24.007  & 41.746    & 9.818    & 31.365  & 52.418    \\ 
\textbf{ComiRec-SA} \cite{cen2020controllable}     & 5.489    & 8.991   & 11.402    & 8.467    & 13.563  & 17.202    & 6.900    & 24.682  & 41.549    & 9.462    & 31.278  & 51.064    \\ \midrule
\textbf{PIMI (Ours)}                                       & \textbf{6.996}    & \textbf{11.221}  & \textbf{14.377}    & \textbf{10.934}   & \textbf{17.094}  & \textbf{21.619}    & \textbf{7.376}  & \textbf{26.003}  & \textbf{43.226}  & \textbf{10.429} & \textbf{33.265} & \textbf{54.043}          \\ \bottomrule
\end{tabular}}
\caption*{Table 2: Performance results on two benchmark datasets (\%). The best
performance in each column is bolded number.}
\label{tab:booktabs}
\end{table*}

In this section, we introduce our experimental setup and 
evaluate the performance of the proposed method, compared 
with several comparable baselines. In order to maintain 
the fairness of comparison, we 
follow the data division and processing method of 
~\citeauthor{cen2020controllable}~\shortcite{cen2020controllable}, 
which are strong generalization conditions. We split all 
users into train/validation/test set according to the 
proportion of 8:1:1, instead of the weak generalization 
condition, where all users are involved in training and 
evaluation. When training, we use the entire sequence 
of the user. Specially, given the behavior sequence 
($i_1^u$, $i_2^u$, ..., $i_k^u$, ..., 
$i_{|{\mathcal{S}^u}|-1}^u$), each training sample 
($i_{k-(n-1)}^u$, ..., $i_{k-1}^u$, $i_k^u$) uses 
the first $n$ items to predict the $(k+1)$-th item, where 
$n$ denotes the maximum sequence length we consider. 
In the evaluation, we take the first 
80$\%$ of the user behavior from validation and test users 
as our model inputs 
to obtain the user's embedding representation, and 
compute metrics by predicting the remaining
20$\%$ user behavior. 
Additionally, we conduct 
a few analysis experiments to prove the effectiveness of PIMI.

\subsection{Experiment Settings}
\paragraph{Datasets.}
We conduct experiments on two publicly available datasets 
\textbf{Amazon}\footnote{http://jmcauley.ucsd.edu/data/amazon/} 
and \textbf{Taobao}\footnote{https://tianchi.aliyun.com/dataset/dataDetail?dataId=649}. 
Amazon dataset includes reviews (rating, text, et al.), 
product metadata (price, brand, et al.), and links from 
Amazon. We use the {\it Books} category of Amazon dataset 
in our experiment. Taobao dataset contains the 
interactive behavior of 1 million users, including  click, 
purchase, adding item to shopping cart and item favoring. We use user click behavior in Taobao dataset for experiment. 
We discard users and items with fewer than 5 interactions, 
and some illegal timestamp information.  
We set the maximum length of training samples for 
Amazon and Taobao to 20 and 50 respectively.
After preprocessing, the statistics of the datasets are 
shown in Table 1.


\paragraph{Baselines.}
To show the effectiveness of the proposed PIMI, we compare 
our model with the following baseline methods: 
(1) \textbf{YouTube DNN} \cite{covington2016deep}  
is a very successful deep learning model for industrial 
recommendation systems, which combines the candidate generation 
model and the ranking model. 
(2) \textbf{GRU4Rec} \cite{hidasi2015session}, which 
is the first work that introduces recurrent neural
networks for the recommendation. 
(3) \textbf{MIND} \cite{li2019multi} is a recent 
model for multi-interest extraction based 
on dynamic routing algorithm. 
(4) \textbf{ComiRec} \cite{cen2020controllable}, which is 
the state-of-the-art model of multi-interest 
extraction, there are two different implementations 
ComiRec-SA and ComiRec-DR based on attention 
mechanism and dynamic routing respectively. 

\paragraph{Evaluation Metrics.}We adopt three common Top-N metrics, Recall@N, NDCG@N 
and Hit Rate@N. Recall@N indicates the proportion of 
the ground truth items are included in the recommendation 
results. 
NDCG@N measures the specific ranking quality
that assigns high scores to hit at top position ranks. 
Hit Rate@N represents the percentage that recommended
items contain at least one ground truth item in top $N$ position.

\paragraph{Implementation Details.}
We implement PIMI with TensorFlow 1.13 in Python 3.7. 
The embedding dimension is 64, batch size for Amazon 
and Taobao are 128 and 256 respectively, dropout 
rate is 0.2, learning rate is 0.001. The time interval 
thresholds for Amazon and Taobao are 64 and 7 
respectively. 
We use three GNN layers to make the items interact sufficiently. 
We set the number of interest embedding is 4, and use 
10 samples for computing sample softmax loss. 
Finally, we iterate at most 1 million rounds in training 
phase. 

\paragraph{The gap between the Training and Testing.}
In the training phase, we select the most relevant user's 
interest embedding for the next target item, 
while in the testing phase, we extract top $N$ items 
for each user’s interest embedding, and then resort 
them according to the value function $Q$. We do this for 
two reasons: (1) Our experiments are conducted with a strong
generalization condition. If the testing phase is consistent
with the training phase, the model only predicts the 
next item based on the most relevant user's interest 
embedding, which is a weak generalization condition 
and not fit real-world situation. (2) For a fair
comparison, we maintain the same experimental conditions
as baselines.

\subsection{Comparisons of Performance}
To demonstrate the sequential recommendation performance of our
model PIMI, we compare it with other state-of-the-art
methods. The experimental results of all methods on
Amazon and Taobao datasets are illustrated in Table 2,
and we have the following observations.

Firstly, YouTube DNN and 
GRU4Rec use a single vector to represent the user, 
while MIND, ComiRec and PIMI use the user’s 
multi-interest representation to make recommendations. 
Experimental results demonstrate that multi-interest 
representation based on user behavior sequence can 
reflect the real-world recommendation situation more adequately. 
Secondly, both MIND and ComiRec-DR 
use dynamic routing method based on capsule network 
to extract multiple interests, while ComiRec-SA 
is a method based on the attention mechanism. We observe 
that on the sparse dataset Amazon, the 
self-attention method can better capture the correlation 
between items in the user behavior sequence, while on 
the dense dataset Taobao, the dynamic routing 
method is better. 
This shows that although self-attention mechanism can 
capture global feature on sparse datasets, the ability 
to capture local feature on dense datasets is 
insufficient. 

Next, our proposed method PIMI consistently outperforms
other competitive methods in terms of three evaluation metrics
on both datasets. 
This indicates that utilizing the timestamp of user 
behavior and encoding time interval information can 
perceive periodicity in multi-interest representation, 
especially the substantial improvement of NDCG metric, 
that means the ranking quality of recommendation 
results is improved due to the addition of time interval 
information, 
which can also demonstrate the effectiveness of the periodicity 
module. 
What's more, experiments show that our interactivity 
module can overcome the problem of long- and short-range 
dependence, allowing the effective interaction of the 
local and non-local features of items, greatly improving 
the performance of the user's multi-interest extraction. 
These results verify the availability and effectiveness 
of PIMI for sequential recommendation.

\begin{table}[t]
\Huge
\resizebox{\linewidth}{!}{
\centering
\begin{tabular}{lcccccc}
\toprule
\multicolumn{1}{c}{\multirow{2}{*}{Metrics@50}} & \multicolumn{3}{c}{Amazon Books} & \multicolumn{3}{c}{Taobao}\\
\multicolumn{1}{c}{}                            & Recall    & NDCG    & Hit Rate  & Recall    & NDCG    & Hit Rate \\ \midrule
PIMI      & \textbf{11.062}    & \textbf{17.228}  & \textbf{21.858}   & \textbf{10.476}    & \textbf{33.502}  & \textbf{54.001}  \\
PIMI-P   & 10.758	& 16.823	& 21.155  & 10.076	& 33.025	& 53.076 \\
PIMI-I  & 9.251	& 14.479	& 18.125 & 9.725	& 32.219	& 50.976 \\ \bottomrule
\end{tabular}}
\caption*{Table 3: Ablation study on two benchmark valid dataset ($\%$).}
\label{tab:booktabs}
\end{table}

\subsection{Model Analysis and Discussion} 
\paragraph{Ablation Study.}
We further investigate that periodicity 
module and interactivity 
module are both essential parts of the 
sequential recommendation task. 
we conduct the ablation studies to compare PIMI with 
PIMI-P and PIMI-I. For the model variant PIMI-P, we 
remove the periodicity module and only make items interact 
via interactivity module. And for the model variant
PIMI-I, we remove the interactivity module and only introduce 
time interval information in periodicity module. 
We show the experimental results of PIMI, PIMI-P, and 
PIMI-I on the Amazon and Taobao valid dataset in Table 3. 
According to the experimental results, we have the 
following observations:
\begin{itemize}
\item PIMI performs better than both PIMI-P nad PIMI-I in 
terms of Recall, NDCG and Hit Rate, which demonstrates 
each component improves the performance effectively. 
\item PIMI-I performs worse than PIMI-P, which indicates 
the effectiveness of our graph structure. The reason for 
this result may be that although items with similar time 
intervals may belong to the same interest, incorrect 
clustering will occur during multi-interest extraction 
without interaction between items.
\end{itemize}

\paragraph{Impact of the virtual central node.}
In order to prove that our graph structure is very 
effective in solving the interactivity between items in
sequence recommendation, we conduct an experiment to 
compare PIMI and PIMI-central\_node. For the
model variant PIMI-central\_node, 
we remove the virtual central node in the graph structure, 
and only model the correlation between adjacent items 
in the sequence. The experimental results in Table 4 
prove that 
only modeling sequential information cannot sufficiently 
explore the interactivity between items. 

\begin{table}[t]
\Huge
\resizebox{\linewidth}{!}{
\begin{tabular}{lcccccc}
\toprule
    & \multicolumn{3}{c}{Metric@20} & \multicolumn{3}{c}{Metric@50} \\ \cmidrule(r){2-4} \cmidrule(r){5-7}
                    & Recall  & NDCG    & Hit Rate  & Recall  & NDCG    & Hit Rate  \\ \midrule
PIMI & \textbf{6.996}   & \textbf{11.221}  & \textbf{14.377}    & \textbf{10.934}   & \textbf{17.094}  & \textbf{21.619} \\
PIMI-central\_node & 6.217   & 10.127  & 13.161    & 9.896   & 15.734  & 20.146    \\ \bottomrule
\end{tabular}}
\caption*{Table 4: Performance comparison for different graph structure 
on Amazon dataset. ($\%$).}
\label{tab:booktabs}
\end{table}

\paragraph{Impact of the time interval threshold.}
Table 5 shows the Metrics@50 of the impact of different 
time interval thresholds on Amazon dataset. 
We choose time interval threshold \{32, 64, 128, 256\} days 
to conduct analysis experiments. 
Experimental results demonstrate that a large 
time interval threshold will lead to sparse 
encodings, and a small time interval threshold will 
cause insufficient learning. The best time interval 
threshold on the Amazon dataset is set to 64. 

\begin{figure*}[h]
\centering
\includegraphics[width=1\textwidth, trim=0 250 0 0, clip]{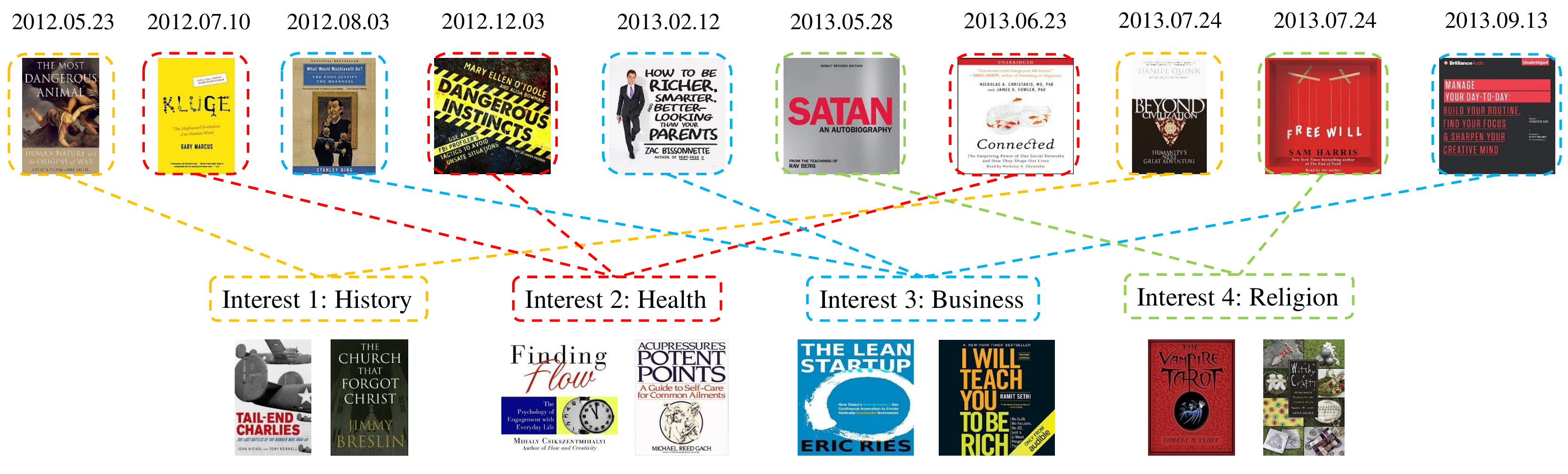}
\caption{A case study of an Amazon user.}
\end{figure*}

\paragraph{Impact of the number of GNN layers.}
\begin{figure}[t]
\centering
\includegraphics[width=0.535\textwidth, trim=70 0 0 0, clip]{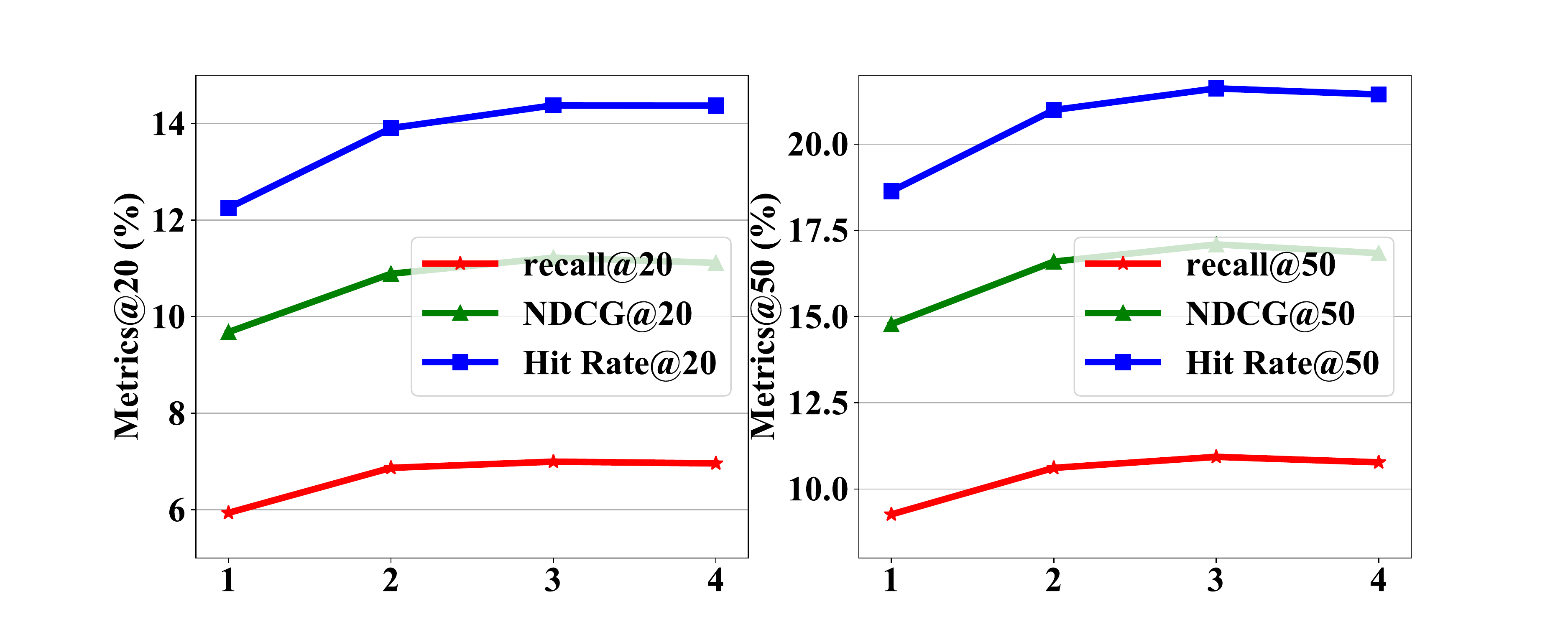}
\caption{Performance comparison for the number of GNN 
layers on Amazon dataset.}
\end{figure}

\begin{table}[t]
\small
\centering
\begin{tabular}{lccc}
\toprule
\multicolumn{1}{c}{\multirow{2}{*}{Metrics@50}} & \multicolumn{3}{c}{Amazon Books} \\
\multicolumn{1}{c}{}                            & Recall    & NDCG    & Hit Rate   \\\midrule
PIMI(threshold=32) & 10.368	& 16.276	& 20.453 \\
PIMI(threshold=64) & \textbf{10.934}  & \textbf{17.094}    & \textbf{21.619} \\
PIMI(threshold=128) & 10.353	& 16.302	& 20.839  \\
PIMI(threshold=256)  & 10.048	& 15.904	& 20.367  \\ \bottomrule
\end{tabular}
\caption*{Table 5: Model performance of Amazon 
dataset for the time thresholds (days)
study in periodicity module ($\%$).}
\label{tab:booktabs}
\end{table}

Figure 4 shows the performance comparison 
for the number of GNN layers on Amazon 
dataset. 
The experimental results demonstrate that as the number 
of layers in GNN increases, the items will learn higher 
quality representation due to the interaction between 
items through $L$ rounds of feature transfer, and the 
performance of our model will be higher. However, 
when the number of GNN layers accumulates to a certain 
extent, the effectiveness of multi-interest extraction 
is slightly reduced due to overfitting, 
and the computational cost will also increase. 

\paragraph{Impact of the number of interests K.}
\begin{figure}[t]
\centering
\includegraphics[width=0.535\textwidth, trim=75 0 0 0, clip]{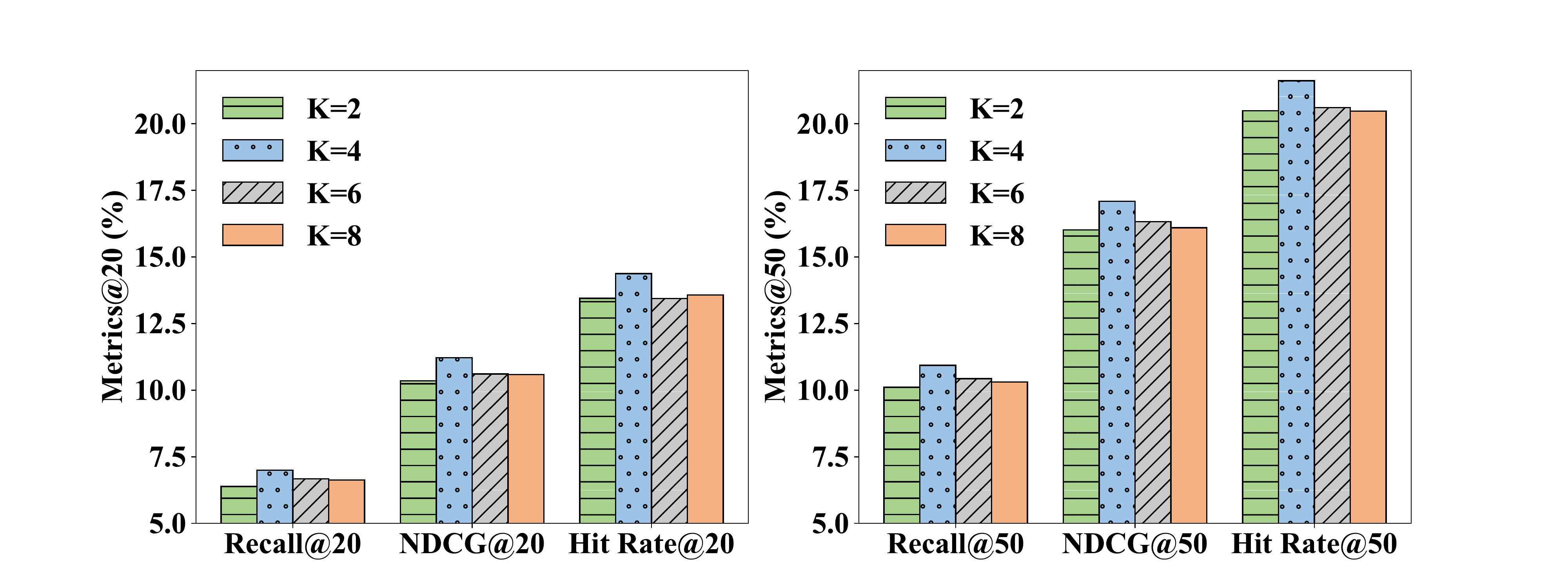}
\caption{Effectiveness of the number $K$ of interests on 
Amazon dataset.}
\end{figure}

Figure 5 shows the Metrics@20 and Metrics@50 of the impact of the number  
$K$ of interests on Amazon dataset. 
For the Amazon dataset, PIMI obtains the 
better performance when K = 4. 
In the real world, the number of interests for each user 
is usually not too much or too little. Hence, 
setting too small and 
too big numbers of interest 
cannot reflect the real situation of users.

\paragraph{Case study.}
As shown in Figure 3, we randomly select a user in the 
Amazon dataset, and generate four interest embedding 
from the user’s behavior sequence. We find that the 
four interests of the user are about 
history, health, business, and religion. 
We have observed that the period for user to review on 
health books is about five months, while the period for 
user to review on business books is about half a year. 
It demonstrates that our proposed PIMI can capture these 
periodicity information successfully, thus contributing  to better representation of interest. 

\section{Conclusion}
In this paper, we proposed a novel method named PIMI for 
sequential recommendation, which shows the effectiveness 
of the periodicity and interactivity of recommendations 
under the multi-interest framework. 
Specifically, we first introduce the periodicity module, 
which constructs the time interval matrix in 
the user behavior sequence, and adds the time 
information to the user's multi-interest representation. 
Next, we design the interactivity module, which captures 
the global and local features of items via a virtual central node, 
and improves the representation quality of items. 
Finally, the multi-interest extraction layer captures 
the user’s multiple interests representation, which can be 
explicitly used to extract candidate items and get  
recommendations. 
Extensive experimental analysis verified that our 
proposed model PIMI consistently outperformed the 
state-of-the-art methods. 
In the future, we plan to explore user modeling issues 
in longer sequence to make better recommendations. 

\section*{Acknowledgments}
This work is supported by National Key Research and 
Development Program of China.

\bibliographystyle{named}
\bibliography{ijcai21}

\end{document}